\documentclass[aps,twocolumn,prd,showpacs,nofootinbib]{revtex4}
\usepackage{amsmath}
\usepackage{graphicx}
\usepackage{amssymb}

\newcommand{\sss}[1]{{\scriptscriptstyle{#1}}}
\newcommand{\be}{\begin{equation}}
\newcommand{\en}{\end{equation}}
\newcommand{\bea}{\begin{eqnarray}}
\newcommand{\ena}{\end{eqnarray}}

\newcommand{\nS}{n_{_{\mathrm{S}}}}

\newcommand{\setR}{\mathbb{R}}
\newcommand{\etal}{\textsl{et al.~}}
\newcommand{\uc}{\mathrm{c}}
\newcommand{\ub}{\mathrm{b}}
\newcommand{\us}{\mathrm{s}}

\newcommand{\uT}{\mathrm{T}}
\newcommand{\uE}{\mathrm{E}}
\newcommand{\usssTE}{\sss{\uT \uE}}
\newcommand{\usssTT}{\sss{\uT \uT}}
\newcommand{\usssEE}{\sss{\uE \uE}}

\newcommand{\zero}{{_0}}
\newcommand{\one}{{_1}}
\newcommand{\two}{{_2}}

\newcommand{\dm}{\mathrm{dm}}

\newcommand{\Mc}{M_\uc}
\newcommand{\OmegaL}{\Omega_\Lambda}
\newcommand{\OmegaCDM}{\Omega_\dm}
\newcommand{\OmegaB}{\Omega_\ub}

\newcommand{\Ps}{P_\mathrm{scalar}}
\newcommand{\ns}{n_\us}
\newcommand{\epsilonT}{\epsilon_\two}
\newcommand{\CAMB}{\texttt{CAMB} }
\newcommand{\COSMOMC}{\texttt{COSMOMC} }
\def\setC{\mathbb{C}}

\def\setR{\mathbb{R}}

\newcommand{\ie}{\textsl{i.e. }}

\newcommand{\GReCO}{${\cal G}\setR\varepsilon\setC{\cal O}$}

\begin{document}

\title{Addendum to ``Superimposed Oscillations in the WMAP Data?''}

\author{J\'er\^ome Martin}
\email{jmartin@iap.fr}
\affiliation{Institut d'Astrophysique de
Paris, \GReCO, FRE 2435-CNRS, 98bis boulevard Arago, 75014 Paris,
France}

\author{Christophe Ringeval}
\email{christophe.ringeval@physics.unige.ch}
\affiliation{D\'epartement de Physique Th\'eorique, Univerist\'e de 
Gen\`eve, 24 quai Ernest Ansermet, 1211 Gen\`eve 4, Switzerland}

\date{\today}

\begin{abstract}
We elaborate further on the possibility that the inflationary
primordial power spectrum contains superimposed oscillations. We study
various effects which could influence the calculation of the multipole
moments in this case. We also present the theoretical predictions for
two other cosmological observables, the matter power spectrum and the
$\mathrm{EE}$ polarization channel.
\end{abstract}

\pacs{98.80.Cq, 98.70.Vc}
\maketitle

\section{Introduction}

The possibility that the multipole moments $C_{\ell }$ characterizing
the angular distribution of the Cosmic Microwave Background Radiation
(CMB) anisotropy on the celestial sphere possess superimposed
oscillations has recently been investigated in Ref.~\cite{MR1}. In
that article, the superimposed oscillations originate from
trans-Planckian wiggles~\cite{tpl} in the primordial inflationary
power spectrum but the study of Ref.~\cite{MR1} was meant to be as
independent as possible from the details of the underlying model. The
presence of oscillations in the primordial spectrum has also been
envisaged in Refs.~\cite{osci}. Then, it has been demonstrated that
the superimposed oscillations can cause a significant improvement of
the fit to the first year Wilkinson Microwave Anisotropy Probe (WMAP) data~\cite{wmap}, thanks to the
presence of the cosmic variance outliers around the first Doppler
peak. Moreover, it has also been shown that the corresponding drop in
the $\chi ^2$ is statistically significant according to the so called
$F$ test. We have since received various inquiries about different
effects that could modify the result obtained in Ref.~\cite{MR1} like
the influence of the splinning, the consideration of the
lensing~\cite{lim} and the way of estimating the statistical
significance of the drop in the $\chi ^2$. In this addendum, we
clarify these issues and, in addition, present new results which are
important for the completeness of Ref.~\cite{MR1}, namely a fit with
an improved $\chi ^2$ which does not suffer from the back reaction
problem and the prediction for two other cosmological observables, the
matter power spectrum and the $\mathrm{EE}$ polarization
channel. Finally, we would like to emphasize that the general
questions analyzed in this Brief Report are important irrespective of
the data set used to study them. Therefore, although we use the first
year WMAP data, the conclusions reached in this addendum are also
valid for the future CMB data releases.

\section{Robustness of the Wiggles}

In this section, we investigate effects that could possibly influence
either the numerical calculation of the CMB multipole moments or the
statistical analysis of the significance of the ``wiggles''.

\subsection{Multipoles splinning}

\begin{figure}
\includegraphics[width=8.6cm]{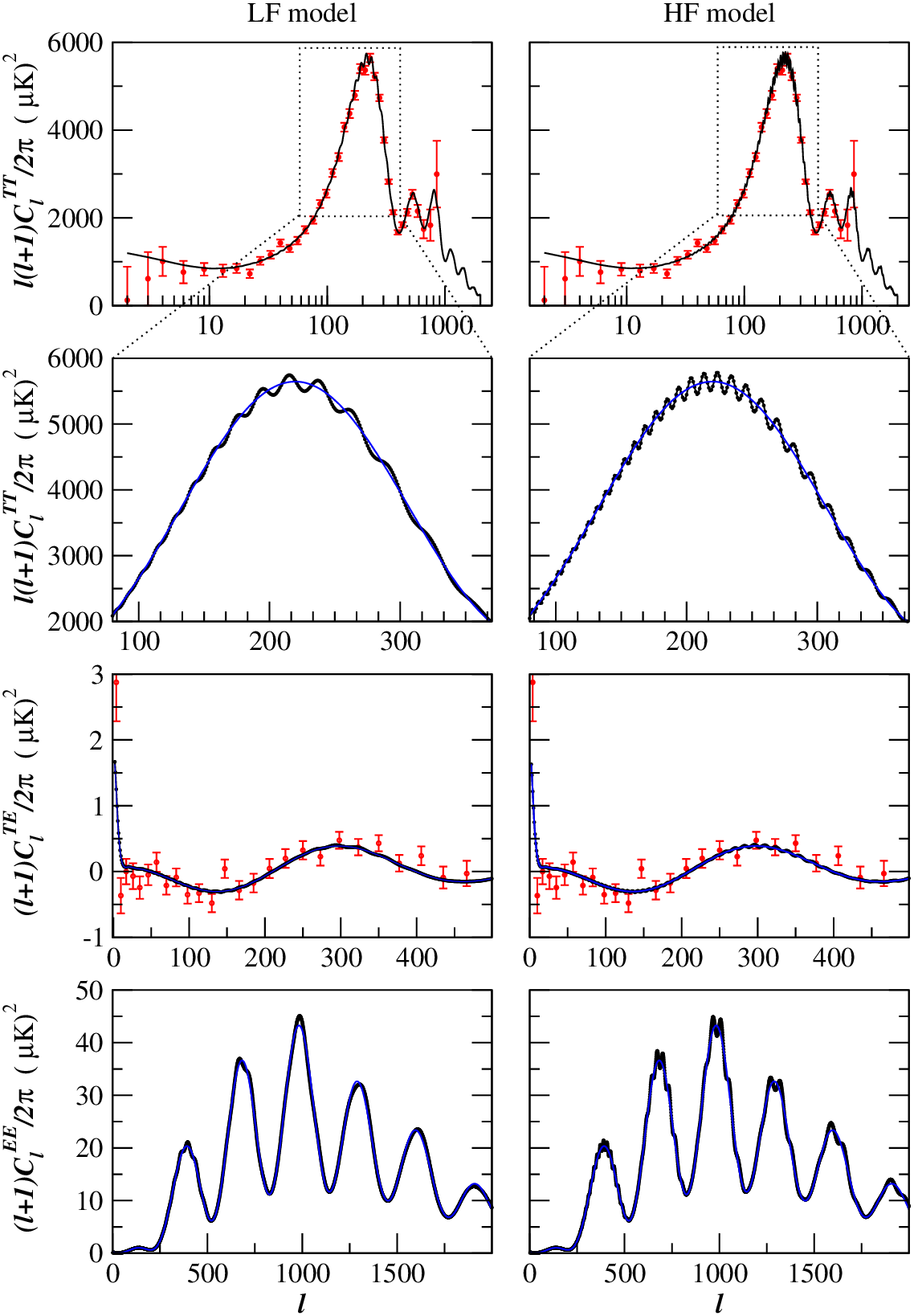}
\caption{The $\mathrm{TT}$, $\mathrm{TE}$ and $\mathrm{EE}$ angular
power spectra for the low frequency (LF) and high frequency (HF)
models. A zoom of the $C_{\ell }^{\usssTT}$'s in the first Doppler
peak region is also shown (black curve) and compared with the standard
slow-roll prediction (blue curve) calculated the with same
cosmological parameters.}
\label{fig:cls}
\end{figure}

In the case where oscillations are present in the initial power
spectrum, a correct numerical computation of the CMB multipole
moments requires one to significantly boost the numerical accuracy of the
code used to derive the so-called transfer
functions~\cite{hu}. Indeed, it is necessary to know these functions
with a high precision if one wants to correctly transfer the
contribution of the primordial oscillating spectrum into each
multipole moments. In Ref.~\cite{MR1}, such high accuracy computations
have been performed with the \CAMB code~\cite{camb} and various tests
have been carried out in order to ensure that the multipole moments
were properly computed. However, one must also pay attention to the
sampling and the splinning performed on the multipole moments. They are
both present by default to avoid prohibitive computation time. The
sampling requires the computation of some multipole moments only and
is scale dependent (\ie $\ell $-dependent) while the splinning
interpolates between those multipole moments and thus allows to
recover the complete angular spectrum. In Ref.~\cite{MR1}, the
sampling and splinning have been kept to their default option which
is clearly not very appropriate to the case where superimposed
oscillations are present. Indeed, as long as the multipole moments
$C_{\ell }$ are not computed for each $\ell $, there is the danger to
undersample the signal which could result in the appearance of
oscillations with an incorrect frequency. At large scales, the
multipole moments are always calculated for each value of $\ell$ and
at very small scales, the sampling has no effect since the
oscillations considered in Ref.~\cite{MR1} are logarithmic in the
Fourier space. Therefore, the danger is particularly present at
intermediate scales, \ie around the first Doppler peak. In this
Brief Report, we have removed the sampling and the splinning from the \CAMB
code such that each multipole is now computed. One finds that this
does not affect  the determination of the multipole moments  as
long as  $\sigma_\zero \gtrsim 4 \times 10^{-4}$ (with
$\varepsilon_\one \simeq 10^{-2}$), but starts to modify the $C_{\ell
}$'s at intermediate scales for smaller values in accordance with the
previous discussion (see Fig~\ref{fig:cls}). Let us also recall that
$\sigma _\zero$ is the dimensionless parameter controlling the 
frequency of the oscillations. It is given by the ratio of the
Hubble parameter during inflation to the scale at which  the new
physics is supposed to show up, $\sigma _\zero \equiv
H/\Mc$. Therefore, we conclude that the sampling or splinning option has
an important effect and must be treated with great care. However,
despite the above discussion, we show below that we essentially
recover the same fit as the one displayed in Ref.~\cite{MR1} up to
some small differences.

\subsection{Lensing}

It has been suggested that CMB lensing effects could blur superimposed
oscillations. This could have an influence on the fit found in
Ref.~\cite{MR1} since this one corresponds to high frequency wiggles.
However, we have checked by mean of the full sky lensing routines
implemented in \CAMB~\cite{camb} that, even in this case, the wiggles
around the first Doppler peak remain almost unchanged. This result is
expected since the damping due to lensing is mainly proportional to
the amplitude of the superimposed oscillations~\cite{hu}. Moreover,
one has to keep in mind that oscillations do not correspond to a
localized feature in the Fourier space as studied in Ref.~\cite{hu},
but, on the contrary, constitute a modification which, in some sense,
is spread everywhere.

\subsection{Exploring the ``fast'' parameter space}

We now turn to the effects that could influence the search of
the likelihood maxima and/or the analysis of their statistical
significance.

\par

As noticed in Ref.~\cite{MR1}, the fact that it is necessary to boost
the computation accuracy in order to correctly transfer the
oscillations significantly increases the computation time for one
model. The use of the sampling (see the discussion above), as in
Ref.~\cite{MR1}, permits us to explore the full (9-dimensional) parameter
space using Monte Carlo methods implemented in the \COSMOMC
code~\cite{cosmomc}. Nevertheless, this exploration remains clearly
limited and may have very well missed the global maximum
of the likelihood function. As a consequence, the fit found in
Ref.~\cite{MR1} should rather been called ``a better fit'' than ``the
best fit''.

\par

In the approach advocated here, where the sampling and
splinning have been removed, the situation becomes even worst. In this
case, an exploration of the full parameter space, even limited, is no
longer possible. In order to tackle this problem, we have fixed the
cosmological parameters to their ``standard values'', determined
from the best fit obtained with a vanilla slow-roll power
spectrum. Then, the parameter space to be explored becomes much
smaller and now consists in the so-called ``fast parameters'' only,
\ie the slow-roll parameters, the frequency and the amplitude of the
oscillations. This method allows to compute the CMB transfer
functions only once and, as a consequence, decreases significantly the
computation time of each Markov chain. But, clearly, one should
keep in mind that, if the actual best fit is somewhere else in the
parameter space, this method will miss it.

\subsection{Comparing the fits}

In Ref.~\cite{MR1}, the ``best fit'' obtained after a limited
exploration of the full parameter space and characterized by
$\chi^2\simeq 1415.4 \, (1340 \, \mathrm{DOF})$ has been compared
to the best inflationary fit published in Ref.~\cite{wmap}, the
$\chi^2 $ of which is given by $\chi ^2\simeq 1431 \, (1342 \,
\mathrm{DOF})$. Hence the number $\Delta \chi ^2\simeq 15$ reported
in that article. However, the best fit of Ref.~\cite{wmap} has been
derived under the assumption that the primordial power spectrum is of
the form $k^{\nS-1}$ which is not exactly the slow-roll
prediction~\cite{LLMS}. In addition, no gravitational waves have been
included whereas this is automatic in the slow-roll approach because
of the consistency check of inflation. Therefore, the number $\Delta
\chi ^2\simeq 15$ does not describe the effect of the wiggles only.
In order to disentangle the influence of the wiggles from the effects
of the other parameters, one should first determine the best fit
obtained with a standard slow-roll power spectrum and then compare
this fit to the one obtained after the addition of the oscillations,
keeping the same values for the cosmological parameters. Our best
slow-roll fit (without the oscillations) corresponds to a model with
$\chi ^2\simeq 1429.3\, (1342\, \mathrm{DOF})$, \ie a difference of
$\Delta \chi ^2\simeq 2$ in comparison with the best fit of
Ref.~\cite{wmap}. Therefore, the effect of adding oscillations is more
accurately described by $\Delta \chi ^2\simeq 13$ rather than $15$. In
the following, the evaluation of $\Delta \chi ^2$ will always refer to
a comparison with this slow-roll fit.

\begin{figure}
\includegraphics[width=8.6cm]{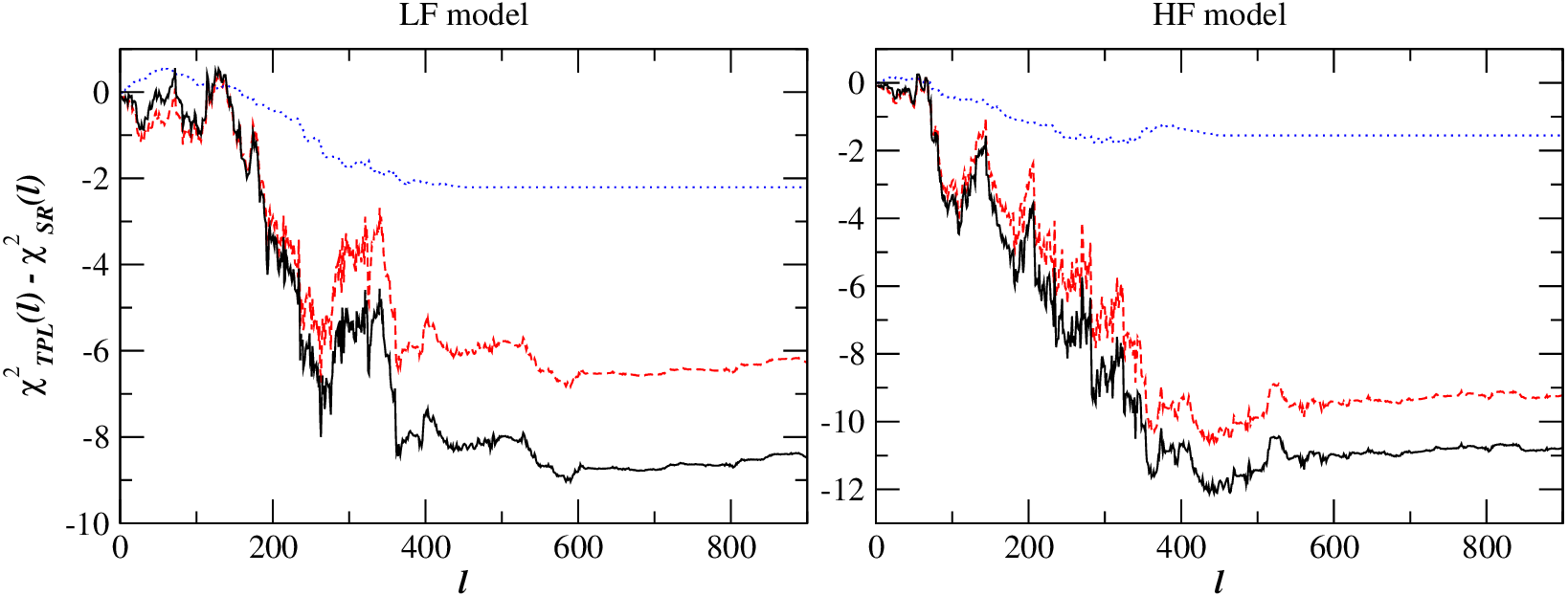}
\caption{Cumulative residual $\chi^2$ for the LF and HF models,
compared to the standard slow-roll one. The dashed red curve
represents the temperature contribution, the dotted blue curve the
polarization contribution and the black curve is the sum of these two
effects.}
\label{fig:deltachi2}
\end{figure}

\subsection{The new improved fits}

\begin{table*}
\begin{tabular}{|c|c|c|c|c|c|c|c|c|c|c|c|c|c|}
\hline Type & $ |x|\sigma_\zero$ & $\sigma_\zero$ & $h$ & $\OmegaB
h^2$ & $\OmegaCDM h^2 $ & $\OmegaL$ & $\tau$& $\Ps$ & $\epsilon_\one $
& $\epsilonT$ & $\ns\equiv 1-2\epsilon _\one -\epsilonT$ &
$\chi^2/\mathrm{DOF}$ \\ \hline HF & $0.107$ & $1.7\times 10^{-4}$
& $0.734$ & $0.024$ & $0.116$ & $0.74$ & $0.1294$ & $23.6 \times
10^{-10}$ & $0.012$ & $-0.0298$ & $1.005$ & $1418.4/1340$\\ LF&
$0.047$ & $3.4 \times 10^{-4}$ & $0.734$ & $0.024$ & $0.116$ & $0.74$ &
$0.1294$ & $23.7\times 10^{-10}$ & $0.010$ & $-0.0226$ & $1.002$ &
$1420.8/1340$\\ \hline
\end{tabular}

\caption{Best HF and LF fit parameters from the WMAP data. The pivot
scale and the time $\eta _\zero$ (see Ref.~\cite{MR1}) are still
chosen such that $k_*/a_\zero =\Mc =0.01 \,\mathrm{Mpc}^{-1}$.}
\label{tbl:bestfit}
\end{table*}

In this section, we present two interesting models corresponding to
two local maxima of the likelihood, the improvements discussed before
having been taken into account. The results are summarized in
Table~\ref{tbl:bestfit}. One model, labeled ``HF'' (high frequency),
corresponds to very rapid oscillations, while the other, labeled
``LF'' (low frequency), has a smaller frequency. The HF model is
essentially similar to the ``best fit'' model found in Ref.~\cite{MR1}
(see the Table I in that article), hence justifying the claim that one
recovers almost the same solution despite the new effects studied in
this addendum. The improvement of the $\chi ^2$ is now $\Delta \chi
^2\simeq 11$ with two additional parameters, that is to say slightly
less good ($\Delta \chi ^2\simeq 2$) but still of the same order of
magnitude as the fit found in Ref.~\cite{MR1}. The corresponding
$F$ probability is $F_{\mathrm{proba}} \simeq 0.6 \%$ and hence the
improvement remains statistically significant. As discussed in
Ref.~\cite{MR1}, this model suffers from a severe back-reaction
problem.

\par

As already mentioned, the LF model seems to be also favored by the
data. Admittedly, the improvement of the $\chi ^2$ is less good with
$\Delta \chi ^2 \simeq 8.5$ for two additional parameters but
corresponds to $F_{\mathrm{proba}}\simeq 1.87\%$, demonstrating that
it is also statistically significant. Furthermore, this model no
longer suffers from the back reaction problem which is, from the
theoretical point of view, an important advantage. Indeed, as can be
seen from Eq.~(11) in Ref.~\cite{MR1}, back reaction effects are
important for small values of $\sigma_\zero$, \ie in the case of high
frequency wiggles. This explains why the LF fit can satisfy the
back reaction constraint.

\par

Finally, in Fig.~\ref{fig:cls}, we present the HF and LF fits and in
Fig.~\ref{fig:deltachi2} we plot the cumulative residual
$\chi^2$ with respect to the standard slow-roll model.

\section{Other cosmological observables}

In Ref.~\cite{MR1}, we have only presented predictions for the
quantities $C_{\ell }^{\usssTT}$ and $C_{\ell }^{\usssTE}$. However,
two other important observables, for which data already exist or will
be available very soon, are the matter power spectrum and the
$\mathrm{EE}$ polarization channel, $C_{\ell }^{\usssEE}$. In this
addendum, we calculate these two observables in the case where
superimposed oscillations are present.

\par

\begin{figure}
\includegraphics[width=8.6cm]{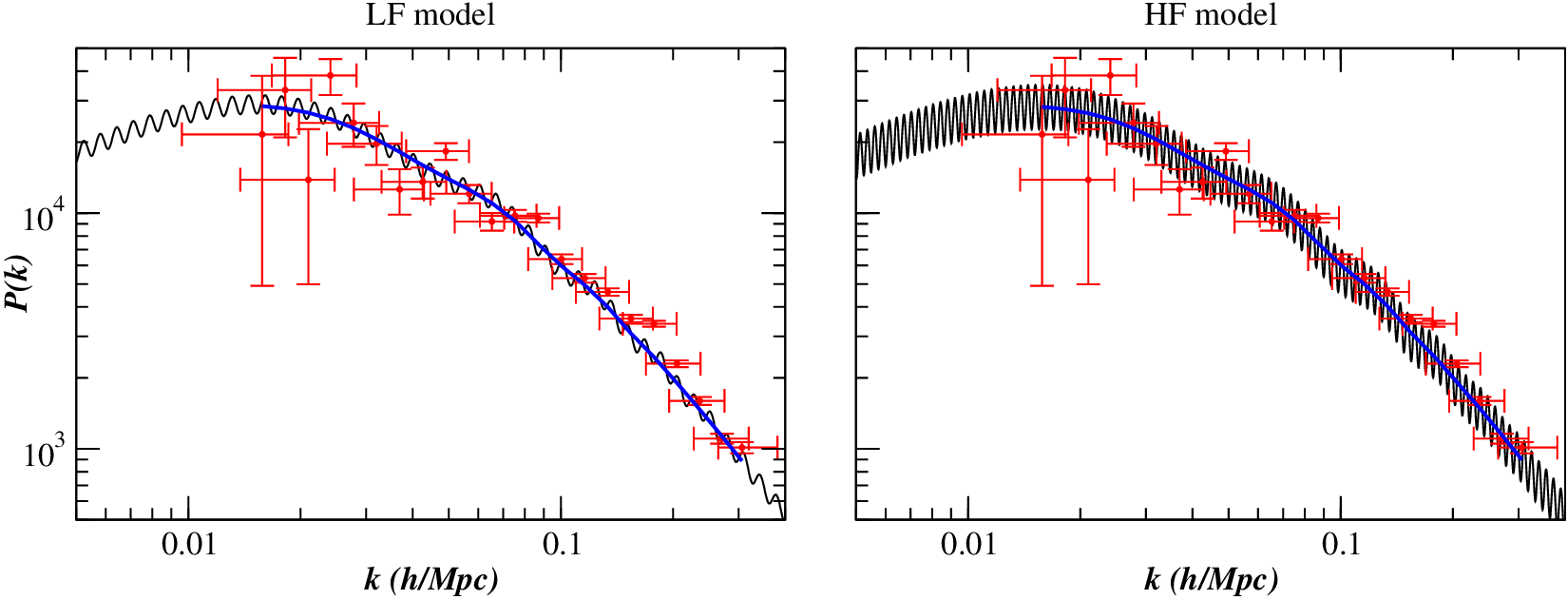}
\caption{Predicted matter power spectra for the LF and HF models
(black curve), compared to the deconvolved SDSS data~\cite{sloan}. The
convolved power spectra by the sloan survey windows function (blue
curve) are also plotted.}
\label{fig:powmat}
\end{figure}

In Fig.~\ref{fig:powmat}, we present the matter power spectrum
corresponding to the two HF and LF best fit values given in
Table~\ref{tbl:bestfit}, together with the current SDSS~\cite{sloan}
deconvolved power spectra in the linear regime. It is obvious that the
oscillations are transfered from the initial power spectrum to the
matter power spectrum since there are linked by a transfer function
only, $P(k)=T(k)P_{\rm \zeta}$. For the HF and LF fits, we see that
the oscillations are well within the error bars. We have also computed
the convolution of the LF and HF oscillatory power spectra with the
sloan survey windows functions (the blue curves in
Fig.~\ref{fig:powmat}). It is clear that the HF and LF convolved
matter power spectra are fully degenerate with the vanilla slow-roll
power spectrum. As a consequence, we conclude that, with the currently
available large scale structure data, no constraint can be put on the
parameters controlling the shape of the oscillations.

\par

Finally, in Fig.~\ref{fig:cls}, we present the $\mathrm{EE}$
polarization multipole moments. As can be seen from the figure, the
oscillations are also transfered to $C_{\ell }^{\usssEE}$. It is
likely that the future polarization measurements will play an
important role in deciding whether the superimposed oscillations are
really present in the data since the drop in the $\chi^2$ is, for the
moment, almost insensitive to polarization (see
Fig.~\ref{fig:deltachi2}). With more accurate polarization data at our
disposal, this situation will certainly change.

\section{Conclusions}

In this addendum, we have investigated various effects that could
influence the presence of wiggles in the CMB multipole moments. We
have shown that the splinning is an important parameter which, in the
presence of high frequency superimposed oscillations, can influence
the shape of the of CMB angular spectrum at intermediate scales. We
have shown that the model presented in Ref.~\cite{MR1} remains favored
by the data (up to a variation of $\Delta \chi ^2\simeq 2$). Two other
cosmological observables have also been presented, namely the matter
power spectrum and the $\mathrm{EE}$ polarization multipole
moments. With the available data, we have concluded that these two
observables do not affect the results reached in Ref.~\cite{MR1}.

\par

A last remark is in order here. In Ref.~\cite{MR1}, the statistical
significance of the wiggles has been discussed by means of the
$F$ test. This was motivated by the fact that the $F$ test only requires
the knowledge of the likelihood maxima for the different models under
consideration. However, by only varying the ``fast parameters'', one
may expect to efficiently probe the corresponding subspace. Therefore,
an interesting improvement of the present work would be to compute the
statistical evidence in this subspace, thus providing us with a
different test of the statistical significance of the wiggles.

\acknowledgments

We are very grateful to Rachel Bean who pointed out to us the fact
that the splinning can affect the shape of the angular spectrum. We
would also like to thank Antony Lewis who suggested to explore the
fast parameter space. We acknowledge useful correspondence with
Olivier Dor\'e, Eugene Lim, Phil Marshall, Hiranya Peiris, Subir
Sarkar, Anze Slozar, Tarun Souradeep and Max Tegmark.


\begin{thebibliography}{99}

\bibitem{MR1} J.~Martin and C.~Ringeval, Phys. Rev. {\bf D69}, 083515
(2004), \eprint{astro-ph/0310382}.
\bibitem{tpl}
J.~Martin and R.~H.~Brandenberger, Phys. Rev. {\bf D63}, 123501
(2001), \eprint{hep-th/0005209}; R.~H.~Brandenberger and J.~Martin,
Mod. Phys. Lett. {\bf A16}, 999 (2001), \eprint{astro-ph/0005432};
J.~Martin and R.~H.~Brandenberger, Phys. Rev. \textbf{D68}, 063513
(2003), \eprint{hep-th/0305161}.
\bibitem{osci} H.~V.~ Peiris \etal Astrophys. J. Suppl. Ser. \textbf{148},
213 (2003), \eprint{astro-ph/0302225}; J.~Barriga, E.~Gaztanaga,
M.~Santos and S. Sarkar, Mon. Not. R. Astron. Soc. \textbf{324}, 977
(2001), \eprint{astro-ph/0011398}; X.~Wang, B.~Feng and M.~Li,
\eprint{astro-ph/0209242}; C.~P.~Burgess, J.~M.~Cline, F.~Lemieux and
R.~Holman, J. High Energy Phys. {\bf 02}, 048 (2003), \eprint{hep-th/0210233};
N.~Kaloper and M.~Kaplinghat, Phys. Rev. \textbf{D68}, 123522 (2003), \eprint{hep-th/0307016}; N.~Kogo,
M.~Matsumiya, M.~Sasaki and J.~Yokoyama, \eprint{astro-ph/0309662};
A.~Shafieloo and T.~Souradeep, \eprint{astro-ph/0312174}; J.~Martin,
P.~Peter and C.~Ringeval, in preparation.
\bibitem{wmap}
D.~N.~Spergel \etal, Astrophys. J. Suppl. Ser. \textbf{148}, 175 (2003),
\eprint{astro-ph/0302209}.
\bibitem{lim}
E.~Lim and T.~Okamoto, Phys. Rev. \textbf{D69}, 083519 (2004), \eprint{astro-ph/0312284}.
\bibitem{hu} W.~Hu, Phys. Rev. {\bf D62}, 043007 (2000),
\eprint{astro-ph/0001303}; W.~Hu and T.~Okamoto, \textit{ibid.} \textbf{69}, 043004 (2004), \eprint{astro-ph/0308049}.
\bibitem{camb}
A.~Lewis, A.~Challinor and A.~Lasenby, Astrophys. J. \textbf{538}, 473
(2000), \eprint{astro-ph/9911177}, \url{http://camb.info}.
\bibitem{cosmomc}
A.~Lewis and S.~Bridle, Phys. Rev. {\bf D66}, 103511 (2002), 
\eprint{astro-ph/0205436}, \url{http://cosmologist.info/cosmomc}.
\bibitem{LLMS}
S.~Leach, A.~R.~Liddle, J.~Martin and D.~J.~Schwarz, Phys. Rev. {\bf 66}, 
023515 (2002), \eprint{astro-ph/0202094}.
\bibitem{sloan}
M.~Tegmark \etal, \eprint{astro-ph/0310725}.


\end{thebibliography}
\end{document}